# Localized surface plasmons selectively coupled to resonant light in tubular microcavities

Yin Yin[1,3], Shilong Li[1,*], Stefan Böttner[1], Feifei Yuan[2], Silvia Giudicatti[1], Ehsan Saei Ghareh Naz[1], Libo Ma[1,*], Oliver G. Schmidt[1,3]

[1]*Institute for Integrative Nanosciences, IFW Dresden, Helmholtzstr. 20, 01069 Dresden, Germany*

[2]*Institute for Metallic Materials, IFW Dresden, Helmholtzstr. 20, 01069 Dresden, Germany*

[3]*Material Systems for Nanoelectronics, Technische Universität Chemnitz, Reichenhainer Str. 70, 09107 Chemnitz, Germany*

*Corresponding authors: S.L.L. (shilong.li.cn@gmail.com) and L.B.M. (l.ma@ifw-dresden.de)

**ABSTRACT**: Vertical gold-nanogaps are created on microtubular cavities to explore the coupling between resonant light supported by the microcavities and surface plasmons localized at the nanogaps. Selective coupling of optical axial modes and localized surface plasmons critically depends on the exact location of the gold-nanogap on the microcavities which is conveniently achieved by rolling-up specially designed thin dielectric films into three dimensional microtube ring resonators. The coupling phenomenon is explained by a modified quasi-potential model based on perturbation theory. Our work reveals the coupling of surface plasmon resonances localized at the nanoscale to optical resonances confined in microtubular cavities at the microscale, implying a promising strategy for the investigation of light-matter interactions.

PACS numbers: 42.60.Da, 73.20.Mf, 52.40.Fd, 78.67.Pt

Noble metal nanostructures are known to support localized plasmonic modes with resonant frequencies in the visible spectral range. These localized surface plasmon (LSP) resonances can be viewed as quasi-static surface plasmon polaritons confined at the metal nanostructures [1, 2], creating intense localized electric fields which can be used for enhanced light-matter interactions [3-6]. Light-matter interactions are very often studied in the context of optical microcavities, where the size mismatch between the optical wavelength and any interacting nano-objects is bridged by cavity quantum electrodynamics [7] or plasmonic nanostructures integrated within the cavities [8-10]. In recent studies, plasmonic nanostructures have been incorporated into microcavities to explore the coupling of surface plasmons confined at metal surfaces and photonic modes supported by dielectric materials [8, 9, 11-16]. For instance, a Fabry-Pérot cavity consisting of two nanostructured metal slabs was fabricated to investigate the coupling of photonic and plasmonic modes by inserting metal nanorods [8, 9]. In addition to Fabry-Pérot cavities, also whispering-gallery-mode (WGM) microcavities have been exploited to observe plasmonic effects. In WGM microcavities, such as microdisks, microtoroids and microcylinders, thin metal layers were deposited on the cavity surface to obtain surface plasmon polaritons or hybrid photon-plasmon resonant modes [13-15]. However, in such cavities there are no distinct localized surface plasmons (LSPs) present due to the flat metal layer coated on the smooth cavity surface. It is therefore of fundamental interest to explore novel plasmonic nanostructures which can efficiently couple to resonant light of optical microcavities.

In this Letter, we demonstrate selective coupling of LSPs confined to metal nanogaps with resonant modes supported by microtubular cavities. Microtubular cavities are obtained by strain induced self-rolling of nanomembranes [17, 18] which have previously received broad interest due to their unique properties including novel optical spin-orbit coupling phenomena [19], injection lasing [20], high opto-fluidic sensitivities [21], and compatibility with on-chip integration technologies [22, 23]. This type of microcavity supports WGM resonances due to self-interference of light propagating along a circular ring trajectory defined by the tube cross section. In addition to the WGMs, the resonant light simultaneously oscillates along the tube axis resulting in the occurrence of well-established axial resonant modes [24-26]. The gold-nanogaps were fabricated on microtubular cavities by depositing a thin gold layer on top of the lobe-patterned spirally rolled-up nanomembranes. By doing so, cavities exhibit nanogap-confined LSPs which can efficiently couple with the resonant light supported by the dielectric microcavities. More significantly, the axial position of this intrinsic vertical nanogap can be tuned by changing the orientation angle of the lobe on the microtubular cavity, allowing for a selective coupling between the optical modes resonating in the lobe area and the LSPs confined to the gold-nanogap.

The microtubular cavities were fabricated by rolling up pre-strained 35 nm thick $SiO_x$ nanomembranes from a U-shaped pattern (see Fig. 1a) [27]. A lobe structure, as the key element in our work, was designed at the middle of the U-shaped pattern. After roll up, an additional 30 nm thick $HfO_2$ layer was grown by atomic-layer-deposition on the microtube surfaces to mechanically strengthen the structure and to optically enhance the light confinement [28]. A scanning electron microscopy (SEM) image of a rolled-up microtube is shown in Fig. 1a. The lobe (before and after rolling) is indicated by the dashed lines. The cross-sectional image of a microtubular cavity is displayed in the left panel of Fig. 1b, acquired by SEM after focused ion beam cutting. A nano-step (~35 nm high, i.e. the thickness of the rolled-up nanomembrane) located at the outer microtube surface is shown in the inset. The nano-step is used as a shadow mask to fabricate gold-nanogaps integrated on the rolled-up microtube cavities, and will be discussed in the following. A simulated WGM mode (with mode number $m = 37$) corresponding to an azimuthal resonance along a ring trajectory of the microtube cross section is shown in the right panel of Fig. 1b. In addition to the azimuthal resonances, axial resonances along the lateral direction (parallel to the tube axis) occur simultaneously due to the axial confinement induced by the lobe structure [24, 27]. Figure 1c shows the optical field distribution in the lobe region for the first four orders of axial modes (i.e. $E_1$ to $E_4$) having the same azimuthal number ($m = 37$).

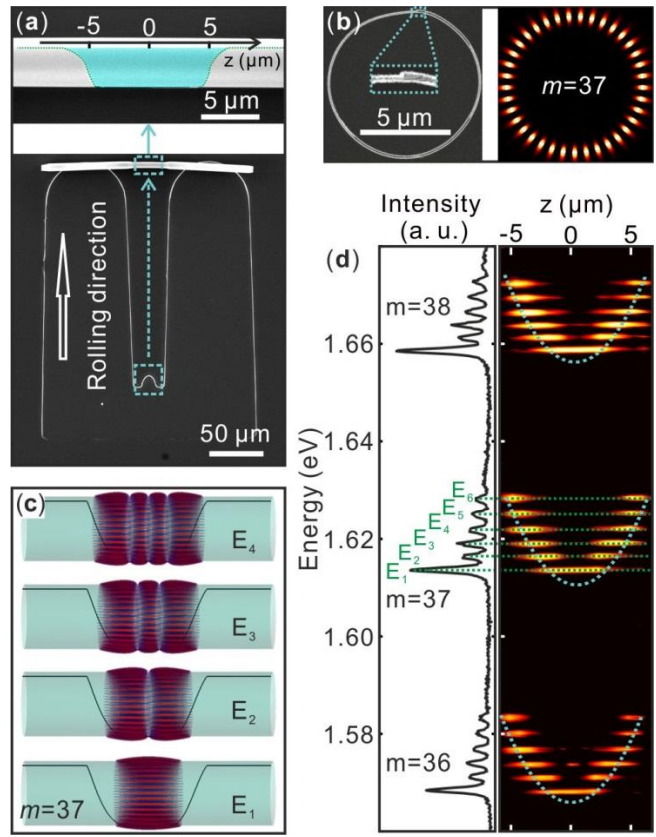

FIG. 1. (color online) (a) SEM images of a rolled-up microtubular cavity. A parabolic lobe structure is located at the middle of the pattern as marked by the dashed square (bottom panel). In the top panel, the lobe after the rolling is indicated by the colored area. (b) Left panel: SEM cross-sectional image of a rolled-up microtubular cavity. The inset shows a nano-step located at the edge of the rolled-up nanomembrane. Right panel: simulated WGM resonance with azimuthal mode number $m = 37$ confined to a tube ring trajectory. (c) Simulated axial modes

with the same azimuthal mode number $m = 37$ but different axial mode numbers (from $E_1$ to $E_4$) which are induced by axial resonances in the lobe region. (d) Right panel: spatial mapping along the cavity axis reveals the distribution of different orders of axial modes. The cyan dotted curves indicate the location of the lobe. The resonant mode spectrum measured at the middle of the lobe ($z = 0$ μm) is shown in the left panel.

Optical resonances in the microcavities were characterized by measuring the light emission with a confocal laser excitation setup. The photoluminescence (PL) of defects [29] in the amorphous silica tube wall was excited by a 442 nm laser line and serves as light source to pump the optical resonances. A 50x objective was used to focus the laser beam onto the tube cavity and collect the light emission from the tube. Figure 1d shows the optical spectrum measured at the center of the lobe region of a $SiO_x$ microtube cavity (left panel). Three groups of resonant modes with azimuthal mode numbers $m$=36-38 are shown in the spectrum, which are associated with azimuthal resonances in the tube ring trajectory. In each group, different orders of axial modes (e.g. E1 to E6) are resolved, which are induced by axial confinement within the lobe region. The axial modes were further revealed by spatial mapping measurements in the lobe region, as shown in the right panel of Fig. 1d. The antinodes, identified by the bright speckles in the mapping figure, denote the optical field distribution of each axial mode within the lobe region. For the higher order axial modes (e.g. $E_5$ and $E_6$), antinodes in the middle area of the lobe are not clearly visible because of limitations in the detection setup.

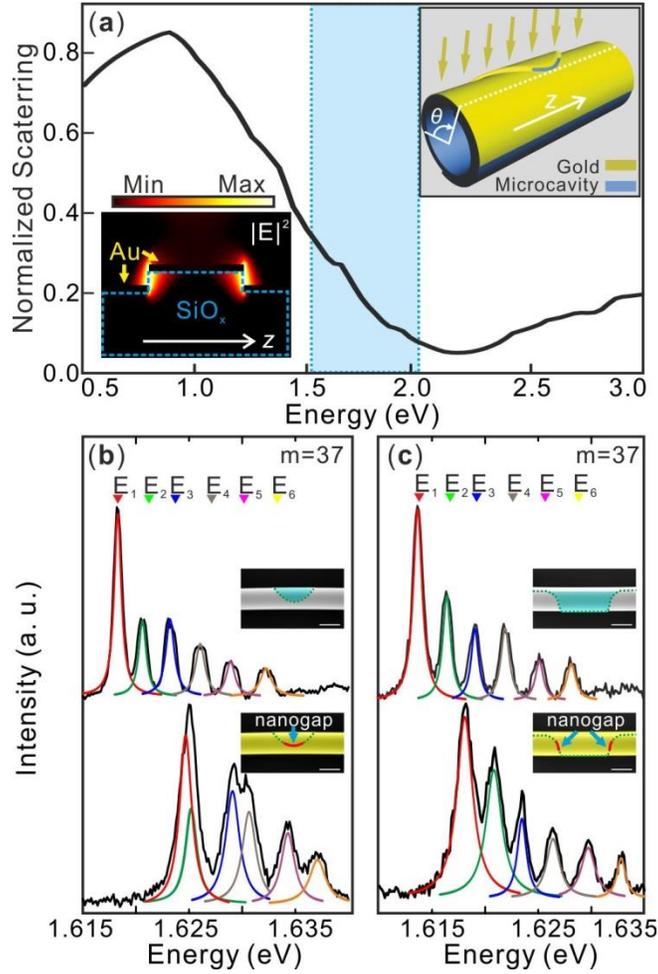

FIG. 2. (color online) (a) Calculated resonant peak of LSPs confined to the gold-nanogap on the cavity surface. The marked region represents the spectral range investigated in the rolled-up microtubular cavities. The upper-right inset shows a sketch of the metal thin film deposited onto a rolled-up microcavity with the lobe orientation defined by the angle $\theta$. The bottom-left inset shows the plasmons resonances supported by the nanogaps located at the lobe edge. (b) and (c) show resonant axial modes supported by two microcavities with lobe orientation angle $\theta\sim90°$ and $\theta\sim240°$, respectively. The top panels present the resonant mode spectra of the bare dielectric cavities, while bottom panels show the mode spectra measured after the deposition of a thin gold film (~10 nm) on top of the microcavities. The black lines correspond to the experimental data, while the colored lines are Lorentzian fits of the individual axial mode $E_1$-$E_6$ identified by the triangular symbols. The insets in (b) and (c) show the colored SEM images of the corresponding rolled-up microcavities, where the locations of the "hot arcs" are marked. The scale bar is 5 μm.

A sketch of the deposited gold layer onto the microtube cavity is shown in the upper-right inset of Fig. 2a. The directional deposition forms a gold layer only on the upper half of the tubular structure, and gold nanogaps can only be present in the top area of the tube, located at the lobe edge. The lobe orientation (and consequently the axial position of the vertical metal nanogaps) can be described by the angle $\theta$ formed by the lobe tip with respect to the horizontal plane. The angle $\theta$ is determined by the rolling length and the tube diameter. LSPs are supported at the metal nanogap via the coupling of the surface plasmons on the metal layer on the lobe surface and the lower metal layer at the lobe foot, respectively [30]. The bottom-left inset of Fig. 2a shows the calculated field distribution of LSPs excited at the vertical nanogaps. Since the vertical nanogaps exist as micro-arcs on the top surface of the tube (see inset of Fig. 2b), these LSPs are called "hot arcs".

The peak of the plasmonic resonance at the vertical nanogap is located at 0.89 eV (1386 nm) which is lower in energy than the optical modes supported by the microtube ranging from 1.55 eV to 2.01 eV (600-800 nm), as marked by the cyan area in Fig. 2a. However, the broad resonant peak of the LSPs overlaps with the spectral range of the studied photonic modes, hence allowing for interaction [31]. The efficiency of the coupling between the resonant axial modes and the "hot arcs" depends on their relative spatial distributions on the tube cavity. When the "hot arcs" spatially overlap with an antinode of the axial mode, a strong interaction occurs leading to a significant spectral shift of the respective axial mode. On the other hand, axial modes without a significant overlap between their antinodes and the "hot arcs" are less affected due to inefficient coupling. Therefore, the optical axial modes supported by the microcavities selectively couple to the LSPs at the "hot arcs" if the mutual position of mode antinode and hot arc coincide.

Two representative microtubular cavities with lobe orientation angles $\theta$ approximately equal to 90° and 240° were used to experimentally investigate the interaction between the optical axial modes and the LSPs at the "hot arcs". The optical modes measured before the gold deposition on the two microcavities are shown in Fig. 2b and c, respectively. As both microcavities have almost the same geometry except for the lobe orientation, they exhibit almost identical resonant spectra. However, a distinct difference of the optical modes is observed after the deposition of a thin gold layer (thickness ~10 nm) on the two microcavities (bottom panels in Fig. 2b and c, respectively). In the microcavity with $\theta$~90°, the lobe tip is located at the center of the top surface of the tube, as shown in the inset of Fig. 2b. In this case, all optical modes experience a slight blueshift while the odd modes (i.e. $E_1$, $E_3$, $E_5$) exhibit an additional pronounced blue-shift. More precisely, the optical modes $E_1$, $E_3$ and $E_5$ experience a blueshift of 6.12, 5.88 and 5.39 meV, respectively, while the even modes only experience a blueshift of 4.65 meV. Moreover, the intensities of odd order axial modes turn out to be significantly larger than even order axial modes after the gold layer deposition, as shown in Fig. 3c. The intensity variation is caused by the efficient coupling between odd order axial modes and the "hot arcs" at the lobe tip. In contrast, all optical modes supported by the second microcavity ($\theta$~240°) show an almost identical blueshift and no relative intensity variation of the axial modes after the gold deposition, as illustrated in Fig. 2c. Upon closer inspection, the higher order modes $E_5$ and $E_6$ show a larger blueshift than the lower order optical modes $E_1$ - $E_4$. This effect will be further discussed in the following part.

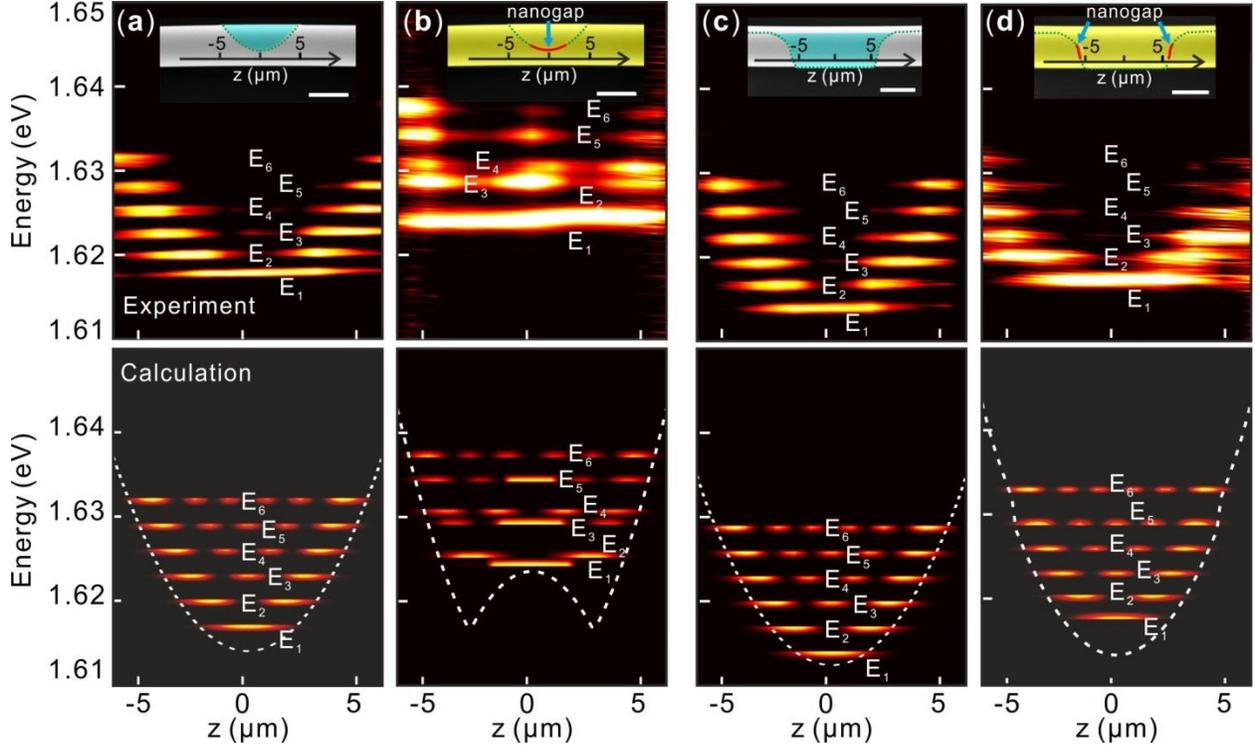

FIG. 3. (color online) Top panels: measured spatial distributions of axial modes in the lobe region for the microcavity with lobe orientation angle $\theta\sim 90°$ before (a) and after (b) gold film deposition, and for the microcavity with lobe orientation angle $\theta\sim 240°$ before (c) and after (d) gold film deposition. Bottom panels: calculated quasi-potential (dashed lines) and antinode distributions of the axial modes for the experimentally investigated microcavities.

The spatial distributions of the axial modes supported by the two microcavities discussed in Fig. 2 were measured by line mappings along the tube axis in the lobe region before and after the deposition of the gold film. These measurements allow for a more detailed investigation of the coupling between optical modes and LSPs. The spectral shifts of the axial modes supported by both microcavities can be clearly observed in the mapping results (see top panels of Fig. 3a-d). For instance, the axial modes of the first microcavity ($\theta\sim 90°$) are equally spaced before the gold deposition (see Fig. 3a), and turn out to be unequally distributed after gold deposition (see Fig. 3b) due to the different shifts experienced by the odd and even order axial modes. Moreover, the faint antinodes in the middle area of the lobe (see Fig. 3a) are clearly visible after gold layer deposition due to the enhancement of the axial modes. This enhancement is caused by the efficient coupling between the axial modes and the "hot arcs" at the lobe tip. For even order axial modes there is no antinode located at the very center of the lobe, therefore their antinodes only partially overlap with the "hot arc" at the lobe tip. Hence, the enhancement of even order axial modes is much weaker than odd order axial modes. For the second microcavity (θ~240°), the "hot arcs" are located at the lobe side-edges (Fig. 3c and d), which can only overlap and interact with the antinodes located at the side edges of the lobe. Thus the antinodes located at the middle of the lobe are not enhanced.

Our observations can be explained by a theoretical model based on an axial potential well modified by the vertical gold-nanogap. In brief, the mode distribution along the axial direction is determined by the lobe structure as described by a quasi-Schrödinger equation [24, 25]: $-\frac{1}{n^2}\frac{\partial^2}{\partial z^2}\psi(z) + k_{circ}^2(z)\psi(z) = k_z^2(z)\psi(z)$, where $n$ is the refractive index, $\Psi(z)$ represents an eigenstate of the axial mode, $k_{circ}(z)$ and $k_z(z)$ denote the quasi-potential and the eigenenergy, respectively. Here, the quasi-potential $k_{circ}(z)$ is determined by solving the optical field $\phi(r,\varphi)$ in the $r$-$\varphi$ plane at each $z$ along the tube axis with the equation $-\frac{1}{n^2}\nabla^2\phi(r,\varphi) = k_{circ}^2(r,\varphi)\phi(r,\varphi)$. By taking into account the geometrical parameters of the two rolled-up microcavities, the quasi-potential, eigenenergies and corresponding eigenstates were calculated, as shown in the bottom panel of Fig. 3a and c. The calculated results agree well with the measured optical modes for both microcavities. The tuning of the optical modes in rolled-up microcavities by post-deposition of thin films has been discussed in previous works [32, 33], and can be incorporated in the equations by modifying the quasi-potential $k_{circ}(z)$ employing perturbation theory [34, 35]. The variation of $k_{circ}(z)$ (=ω/c) caused by the thin film deposition is determined by

$$\Delta\omega = -\frac{\omega}{2}\frac{\langle E(r,\varphi)|\Delta\varepsilon(r,\varphi)|E(r,\varphi)\rangle}{\langle E(r,\varphi)|\varepsilon(r,\varphi)|E(r,\varphi)\rangle}$$, where $\omega$, $E(r,\varphi)$, and $\varepsilon(r,\varphi)$ are the angular frequency, the azimuthal optical field, and the permittivity, respectively. $\Delta\varepsilon(r,\varphi)$ denotes the variation of the permittivity induced by the presence of the thin metal film. Considering the LSPs confined in the vertical nanogap at the lobe edge, the electric field is enhanced with an intensity of $|E_{LSP}|^2$ in excess of $|E_{r,\varphi}|^2$ when spatially overlapping with the LSPs, which in turn modifies the axial potential [36-38]. The measured axial mode distribution and shift agree well with the results of the calculations in the modified quasi-potential wells, as shown in the bottom panel of Fig. 3b and d. After the metal film deposition on the first microcavity ($\theta\sim90°$), the bottom of the potential well is reversed, leading to significant energy shifts of the odd order modes. This effect is explained by the fact that a centered antinode of the odd order modes is strongly influenced by the reversed potential, as shown in Fig. 3b. For the second microtube cavity ($\theta\sim240°$), the potential was slightly modified at the two top-side edges where the LSPs interact with the higher order optical modes ($E_5$ and $E_6$), leading to a slightly increased mode shift for $E_5$ and $E_6$.

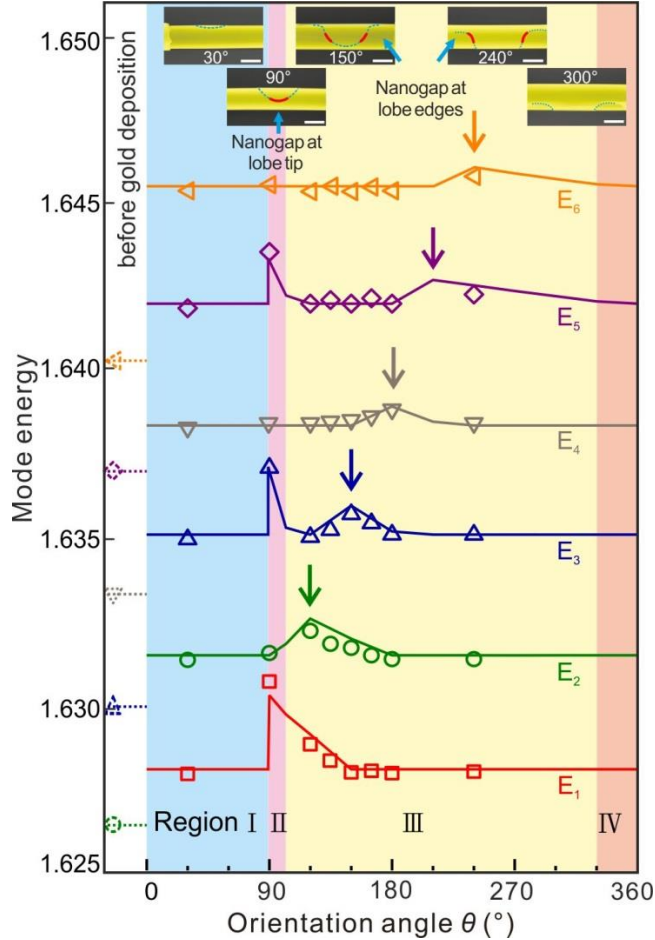

FIG. 4. (color online) Calculated (solid lines) and measured (symbols) energy of the resonant axial modes from $E_1$ to $E_6$ after depositing the gold layer (~10 nm thick) on top of the microcavities having different lobe orientation angle $\theta$. To compare the relative mode shifts, all measured resonant modes of each tube were calibrated to have the same initial position. The error bar matches the symbol size.

The good agreement between systematic measurements (see Supplemental Material) and theoretical calculations are shown in Fig. 4, where the "hot arcs" are located at different positions on the tube surface. After the gold deposition, all optical modes blueshift because of the negative permittivity of gold [39]. When $\theta$ increases from 0 to 90°, the vertical deposition of the gold layer does not result in a metal nanogap at the lobe, as sketched in Region I of Fig. 4. In this region, LSPs are not present and the simultaneous shift of all optical modes is purely determined by the thickness of the gold layer. When $\theta$ exceeds 90° (Region II), a vertical metal nanogap that supports LSPs appears near the lobe tip. In this case, the odd order modes $E_1$, $E_3$ and $E_5$ exhibit a larger blueshift in comparison to the even order axial modes $E_2$, $E_4$ and $E_6$. This is due to the fact that the central antinode of the odd order modes overlaps with the "hot arc" located at the lobe tip, allowing for an efficient coupling between the LSPs and the odd order modes. The even order modes, instead, do not possess an antinode in the central part of the lobe which could couple to the LSPs at the lobe tip. For $\theta$ larger than 100° (Region III), the "hot arc" splits into two components symmetrically located at the two side edges of the lobe. In this region, the "hot arcs" selectively couple to different order axial modes depending on the locations of "hot arcs" at the lobe edge. The axial modes exhibit a significant spectral shift when coupled with "hot arcs", as indicated by the arrows in Fig. 4.

In addition, one can see that the energy shifts of the coupled higher order axial modes decrease as $\theta$ increases in the range of 90-330°. This is explained by the fact that the higher order axial modes exhibit a weaker field strength overlapping with the "hot arcs", leading to a smaller mode shift. For $\theta$ corresponding to region IV (with the lobe structure rolled to the down side of the cavity and consequently no formation of nanogaps during the metal deposition), the energy shifts of the modes are similar and remain constant because of the absence of the LSPs. For the lobe orientation angle $\theta$ larger than 330° (Region IV), the lobe structure is shielded by the tube from the directional gold deposition so that no nanogaps are formed, again resulting in only a uniform blueshift of the axial modes.

In summary, we demonstrated efficient coupling between optical resonant modes supported by microtubular cavities and surface plasmons localized at vertical gold-nanogaps. The plasmonic nanogap was fabricated by depositing a thin gold layer onto the nano-step of a lobe-patterned rolled-up microtubular cavity. The optical axial modes confined within the lobe region can selectively couple to the localized surface plasmons depending on the location of the vertical gold-nanogaps around the lobe profile. This selective coupling between optical axial modes and localized surface plasmons is explained by a modified quasi-potential model based on perturbation theory. Our work reveals the interaction between surface plasmon resonances localized at the nanoscale and optical resonances confined in WGM microcavities at the microscale, thus establishing a unique platform for future investigations of light-matter interactions.

The authors thank R. Engelhard, S. Harazim, B. Eichler and S. Baunack for technical support. This work was supported by the Volkswagen Foundation (I/84072) and the DFG research group No. FOR 1713. Y.Y. acknowledges support by China Scholarship Council under file No. 201206090008. S. Li acknowledges support by China Scholarship Council under file No. 2008617109.